\documentclass[twocolumn,preprintnumbers,superscriptaddress,nofootinbib,aps,prd,floatfix]{revtex4}

\usepackage{enumerate}
\usepackage{subfigure}
\usepackage{amsmath}
\usepackage{amssymb}
\usepackage{graphicx}
\usepackage{xspace,slashed}
\usepackage{hyperref}
\hypersetup{colorlinks=true, citecolor=blue, urlcolor=blue, linkcolor=blue}
\usepackage{bbm}

\begin{document}

\providecommand{\abs}[1]{\lvert#1\rvert}

\preprint{DESY-23-082}

\title{Higgs Footprints of Hefty ALPs}
\begin{abstract}
We discuss axion-like particles (ALPs) within the framework of Higgs Effective Field Theory, targeting
instances of close alignment of ALP physics with a custodial singlet character of the Higgs boson. 
We tension constraints arising from new contributions to Higgs boson decays against
limits from high-momentum transfer processes that become under increasing control 
at the LHC. Going beyond leading-order approximations, we highlight the importance 
of multi-top and multi-Higgs production for the pursuit of searches for physics beyond 
the Standard Model extensions.
\end{abstract}

\author{Anisha} \email{anisha@glasgow.ac.uk}
\affiliation{School of Physics and Astronomy, University of Glasgow, Glasgow G12 8QQ, United Kingdom\\[0.1cm]}
\author{Supratim Das Bakshi}\email{sdb@ugr.es}
\affiliation{CAFPE and Departamento de F\'isica Te\'orica y del Cosmos, Universidad de Granada,
Campus de Fuentenueva, E–18071 Granada, Spain\\[0.1cm]}
\author{Christoph Englert} \email{christoph.englert@glasgow.ac.uk}
\affiliation{School of Physics and Astronomy, University of Glasgow, Glasgow G12 8QQ, United Kingdom\\[0.1cm]}
\author{Panagiotis Stylianou} \email{panagiotis.stylianou@desy.de}
\affiliation{Deutsches Elektronen-Synchrotron DESY, Notkestr. 85, 22607 Hamburg, Germany\\[0.1cm]}

\maketitle
\allowdisplaybreaks
\section{Introduction}
\label{sec:intro}
Searches for new interactions beyond the Standard Model of Particle Physics have, so far, been unsuccessful. This is
puzzling as the Standard Model contains a plethora of flaws that are expected to be addressed by a more comprehensive
theory of microscopic interactions. A reconciliation of these flaws can have direct phenomenological consequences for physics at or below 
the weak scale $v\simeq 246~\text{GeV}$. This is particularly highlighted by fine-tuning problems related to the Higgs mass or
the neutron electric dipole moment, both of which take small values due to cancellations which are not protected by symmetries in
the Standard Model (SM). Dynamical solutions to these issues have a long history, leading to new interactions and states around the TeV scale
to address Higgs naturalness, or relaxing into and CP-conserving QCD vacuum via a Peccei-Quinn-like mechanism~\cite{Peccei:1977hh}. 
Often such approaches yield an additional light pseudo-Nambu Goldstone field, in the guise of a composite Higgs boson or the axion~\cite{Wilczek:1977pj}.

The search for a wider class of the latter states referred to as axion-like particles (ALPs) bridges different areas of high 
energy physics. Efforts to detect ALPs across different mass and coupling regimes have shaped the current BSM programme in many
experimental realms (see e.g.~\cite{Graham:2015ouw,Irastorza:2018dyq} for recent reviews). In particular, at the Large Hadron Collider (LHC), ALP interactions
have been discussed in relation to their tell-tale signatures arising from $\sim F\tilde F$ coupling structures~\cite{Jaeckel:2015jla}, top quarks~\cite{Esser:2023fdo},
emerging signatures~\cite{Alonso-Alvarez:2023wni}, flavour physics~\cite{Bauer:2021wjo,Bauer:2021mvw}, electroweak precision constraints~\cite{Bauer:2017ris}, Higgs decays~\cite{Bauer:2017ris,Biekotter:2022ovp}, and mixing~\cite{Bauer:2022rwf}. The methods of
effective field theory~\cite{Bonilla:2021ufe} naturally embed ALP-related field theories into a broader framework of a more modern perspective
on renormalisability~\cite{Buchalla:2013rka,Buchalla:2017jlu,Buchalla:2020kdh}.

Experimental searches for these states have been carried out using a variety of techniques, including collider searches, precision measurements of atomic and nuclear transitions (e.g. ACME~\cite{PMID:30333583} and nEDM~\cite{PhysRevLett.124.081803}), and searches from astrophysical events~\cite{Raffelt:1996wa,PhysRevD.37.1237}, over a wide range of ALP mass \cite{AxionLimits}. In particular for the ALP mass range $ M_\mathcal{A} \, \in $ \, [6, 100] GeV, the most stringent exclusion limits for ALPs are derived from ultra-peripheral lead nuclei collision data~\cite{ATLAS:2020hii,CMS:2018erd}. These limits are from exclusive di-photon searches, and define SM--ALPs interactions via electromagnetic interactions $(\sim F \widetilde{F})$. \footnote{When the ALP mass $ M_\mathcal{A} < 2 m_e $, only the di-photon channel is the allowed decay process via SM particles. In same manner, with greater ALP mass, the decay modes to other leptons, quarks (jets), gauge bosons open up as well.} The ATLAS limits~\cite{ATLAS:2020hii} on the ALP--photon cross sections when put in terms of ALP--photon couplings is found in the range $ g_{\mathcal{A}\gamma} \in [0.05,1]$ TeV$ ^{-1} $ \cite{dEnterria:2021ljz}. However, in general, the ALP--SM interactions can be defined via gauge bosons, fermions, and scalars; although, its decays will depend on its mass. The limits on ALP couplings to the SM fields (except photons) are less stringent. The exotic decays of the SM Higgs and Z-bosons are promising channels for ALP searches (particularly benefiting from the high-luminosity run of the LHC), e.g., with the decay modes $ h \rightarrow Z \mathcal{A}$~\cite{CMS-PAS-HIG-22-003}.

It is the latter perspective that we adopt in this note to focus on ALP interactions with the Higgs boson, also beyond leading order.
Adopting the methodology of Higgs Effective Theory (HEFT),
we can isolate particular interactions of the ALP state and trace their importance (and thus the potential for constraints) to representative collider processes
that navigate between the low energy precision and the large momentum transfer regions accessible at the LHC. If the interactions of ALPs and Higgs particles
is predominantly related to a custodial singlet realisation of the Higgs boson, these areas might well be the first phenomenological environments where
BSM could be unveiled as pointed out in, e.g.,~Ref.~\cite{Brivio:2017ije}. In parallel, our results demonstrate the further importance of multi-top and multi-Higgs
final states as promising candidates for the discovery of new physics. With the LHC experiments closing in on both Higgs pair~\cite{CMS:2022hgz,ATLAS:2023qzf} and four-top production in the SM~\cite{CMS:2023ftu,ATLAS:2023ajo}, such searches becoming increasingly interesting for our better understanding of the BSM landscape.

This work is organised as follows: In Sec.~\ref{sec:alpheftlag},
we review the ALP-HEFT framework that we use in this study to make this work self-contained (a comprehensive discussion is presented in~\cite{Brivio:2017ije}). In Sec.~\ref{sec:hzax}, we focus on the decay phenomenology of the Higgs boson in the presence of ALP interactions before we turn to discuss a priori sensitive processes that can provide additional constraints due to their multi-scale nature and kinematic coverage. Specifically, in Sec.~\ref{sec:virt} we analyse ALP corrections to Higgs propagation as accessible in four-top final states~\cite{Englert:2019zmt,CMS:2019rvj,ATLAS:2023ajo}, which informs corrections to multi-Higgs production.
We conclude in Sec.~\ref{sec:conc}.

\section{ALP Chiral HEFT Lagrangian}
\label{sec:alpheftlag}
The leading order ALP interactions with SM fields in the framework of chiral (non-linear) electroweak theory are written as
\begin{equation}
	\label{eq:lag}
\mathcal{L}_{\text{LO}}=\mathcal{L}^{\text{HEFT}}_{\text{LO}}+ \mathcal{L}^{\text{ALP}}_{\text{LO}}\,.
\end{equation}
$\mathcal{L}^{\text{HEFT}}_{\text{LO}}$ is the chiral dimension-2 HEFT Lagrangian~\cite{Alonso:2012px,Brivio:2016fzo,Herrero:2020dtv,Herrero:2021iqt,Anisha:2022ctm}. In this framework, the SM Higgs ($H$) is a singlet field and the Goldstone bosons $\pi^{a}$ are parametrised non-linearly using the matrix $U$
\begin{equation}
	U(\pi^{a}) = \exp\left({i \pi^{a} \tau^{a}/v}\right)\,,
\end{equation}
with $\tau^{a}$ as the Pauli matrices with $a= 1,2,3$ and $v\simeq 246~\text{GeV}$. The $U$ matrix transforms under $L\in SU(2)_L, U(1)_Y\subset SU(2)_R \ni R $ as $U\to L U R^\dagger$ and is expanded as
\begin{equation}
	U(\pi^{a}) =  {\mathbbm{1}}_{2} + i \frac{\pi^{a}}{v} \tau^{a} - \frac{2G^{+}G^{-} + G^{0} G^{0}}{2 v^2}  {\mathbbm{1}}_{2} + \dots \,,
\end{equation}
where $ G^{\pm} = ( \pi^{2} \pm i \pi^{1} )/\sqrt{2}$ and $ G^{0} = -\pi^{3}$. The dynamics of the gauge bosons $W^{a}_{\mu}$ and $B_{\mu}$ are determined by the usual $SU(2)_{L} \times U(1)_{Y}$ gauge symmetry. Weak gauging of $SU(2)_L\times U(1)_Y$ is achieved through the standard covariant derivative
\begin{equation}
	D_{\mu}U = \partial_{\mu} U + i g_{W} (W^{a}_{\mu} \tau^{a} /2) \; U -i g' U B_{\mu}\tau^{3}/2 \,.
\end{equation}
The gauge fields in the physical (mass and electromagnetic $U(1)_\text{em}$) basis are are related to the gauge basis via the Weinberg angle $s_W=\sin\theta_W,c_W=\cos\theta_W$
\begin{equation}
\begin{split}
	W^{\pm}_{\mu} &=   \frac{1}{\sqrt{2}}(W^{1}_{\mu} \mp W^{2}_{\mu})\,,\\
	\begin{pmatrix}
		Z_{\mu} \\ A_{\mu}
	\end{pmatrix} &= \begin{pmatrix}
		c_{W} & s_{W} \\
		-s_{W} & c_{W}
	\end{pmatrix}
	\begin{pmatrix}
		W^{3}_{\mu} \\ B_{\mu}
	\end{pmatrix}\,.
\end{split}
\end{equation} 
The leading order HEFT Lagrangian relevant for our discussion is then given by
\begin{subequations}
	\label{eq:heftlo}
	\begin{multline}
		\mathcal{L}^{\text{HEFT}}_{\text{LO}} = - \frac{1}{4} W^{a}_{\mu \nu} W^{a\mu \nu} -  \frac{1}{4} B_{\mu \nu} B^{\mu \nu} + {\cal{L}}_{\text{ferm}}+ \mathcal{L}_{\text{Yuk}} \\+ \frac{ v^2}{4} \mathcal{F}_{H} \,\text{Tr}[D_{\mu} U^{\dagger} D^{\mu} U]   \\
		+ \frac{1}{2} \partial_{\mu} H \partial^{\mu} H - V(H) \,.
	\end{multline}
The interactions of the singlet Higgs field with gauge and Goldstone bosons are parametrised by the flare function $\mathcal{F}_{H}$ given as
\begin{equation}
	 \label{eq:flare}
		\mathcal{F}_{H} = \Big(1+ 2(1+\zeta_{1})\frac{H}{v} + 
		 (1+\zeta_{2}) \Big(\frac{H}{v}\Big)^2 + 
		... \Big)\,.
	\end{equation}
\end{subequations}
The couplings $\zeta_{i} $ denote the independent parameters that determine the leading-order interactions of the Higgs boson with the gauge fields. ${\cal{L}}_{\text{ferm}}$ parametrises the fermion-gauge boson interactions, which we take SM-like in the following. $V({H})$ is the Higgs potential, which we relate to the SM expectation
\begin{equation}
\label{eq:hpot}
	V({H})=  \frac{1}{2} M_{H}^2 H^2 + \kappa_3
	H^3+ \kappa_4 
	H^4 \,.
\end{equation}
with $\kappa_3\simeq 32~\text{GeV}, \kappa_4\simeq 0.03$ in the SM.

In this work, we consider ALP interactions that particularly probe the singlet character of the Higgs boson as a parametrised by the HEFT Lagrangian. The interactions are given by
\begin{subequations}
\label{eq:axlag}
\begin{multline}
 \mathcal{L}^{\text{ALP}}_{\text{LO}} =  \frac{1}{2} \partial_{\mu} \mathcal{A} \partial^{\mu} \mathcal{A} -\frac{1}{2}M_{\mathcal{A}}^{2} \mathcal{A}^2 \\ 
 + a_{2D} \Big(i \, v^2 \text{Tr}[ U \tau^3 U^{\dagger} \,	\boldsymbol{\mathcal{V}}_{\mu}]  \; \frac{\partial_{\mu} \mathcal{A}}{f_{A}} {\mathcal{F}}_{2D}\Big)\,
\end{multline}
with
\begin{equation}
 \label{eq:alpflare}
	{\mathcal{F}}_{2D}=  \Big(1+ 2\zeta_{12D}\frac{H}{v} + \zeta_{22D} \Big(\frac{H}{v}\Big)^2 + ... \Big)\,,
\end{equation}
and
\begin{equation}
	\boldsymbol{\mathcal{V}}_{\mu}= (D_{\mu}U)U^{\dagger}\,.
\end{equation}
\end{subequations}
In Eq.~\eqref{eq:axlag}, $f_{\mathcal{A}}$ denotes the scale linked with the ALP interactions. The interactions specified by Eq.~\eqref{eq:axlag} are the leading order chiral
interactions of the ALP field with SM states. These couplings specifically probe the custodial singlet nature of the Higgs boson~\cite{Brivio:2017ije}. Therefore, the phenomenology of these interactions provides relevant insights into the mechanism of electroweak symmetry breaking and its relation to axion-like states. The radiative imprints of these interactions on SM correlations are then captured by the chiral dimension-4 interactions contributing to
$\mathcal{L}^{\text{HEFT}}_{\text{LO}}$ when HEFT parameters coincide with the SM expectation.\footnote{All interactions detailed above are implemented using the {\sc{FeynRules}} package~\cite{Christensen:2008py,Alloul:2013bka}.}
  
The aim of our analysis is to clarify the phenomenological reach to the couplings involved in Eq.~\eqref{eq:axlag} from two different angles. Firstly, these interactions
are clear indicators of a singlet character of the Higgs boson in HEFT. Secondly, the
interactions $\sim\zeta_{12D}$ will introduce modifications to the Higgs boson propagation and Higgs decay in HEFT, $\zeta_{22D}$ will imply
modifications to the Higgs pair production rate. Although the ALP might be too light to be directly accessible 
at collider experiments such as the LHC, its virtual imprint through specific predictions between the correlations of four-top and Higgs pair
production could reveal its presence. We will turn to the expected constraints in the next section. Throughout, we will identify the HEFT parameters with their corresponding tree-level SM limit except for the deviations introduced by the ALP, which we also detail below. We will focus on the interactions that are generated at order $a_{2D}\zeta_{12D}$ etc.; fits against the ALP-less HEFT (or the SM as a particular HEFT parameter choice) should be sensitive to these contributions when data is consistent with the latter expectation. To reflect this we will therefore also assume that HEFT operators coincide with their SM expectation. Specifically this means that we will choose vanishing HEFT parameters arising at chiral dimension-4. Departures from the SM correlations are then directly related to (radiative) presence of the ALP.

\section{Direct constraints from Higgs decays}
\label{sec:hzax}
The interactions of an ALP with a Higgs boson via the $a_{2D}\zeta_{12D}$ coupling of Eq.~\eqref{eq:axlag} is tree-level mediated. The exotic decay of the Higgs boson via $H  \rightarrow \mathcal{A}  \, Z$ at leading order is given by 
\begin{equation}
\label{eq:dec}
	\parbox{2.7cm}{\includegraphics[width=2.7cm]{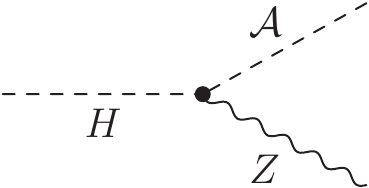}}
	=- 2i \frac{e}{c_W s_W} \frac{ v  }{ f_{\mathcal{A}} } a_{2 D} \,\zeta _{12 D}\, q^{\mu}(H)\,,
\end{equation}
with $q^\mu(H)$ denoting the four-momentum carried by the Higgs leg. When kinematically accessible, the decay width of the Higgs boson 
receives a non-SM contribution
\begin{multline} \label{eq:dwhaz}
	\Gamma( H \rightarrow \mathcal{A} Z) =   \frac{  v^2 a_{2D}^2 \zeta_{12D}^2}{274 \; s_{W}^2 c_{W}^2 f_{\mathcal{A}}^2  M_{H}^3  M_{Z}^2 }\\
	\Big(M_{\mathcal{A}}^4-2 M_{\mathcal{A}}^2 (M_H^2+M_{Z}^2)+(M_{H}^2-M_{Z}^2)^2\Big)^{3/2}.
\end{multline}
Assuming this two-body process as the most dominating BSM decay involving the ALP, the SM Higgs boson signal strengths get uniformly
modified
\begin{figure}[!t]
	\includegraphics[width=0.45\textwidth]{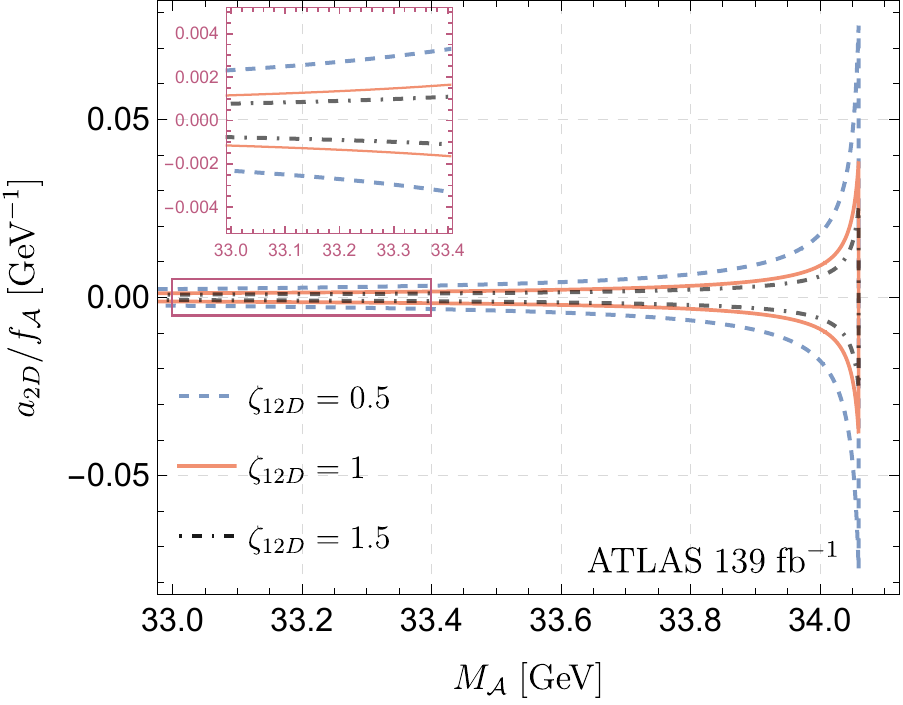}
	\caption{ \label{fig:a2Dvsma}The allowed parameter region obtained from the di-photon signal strength reported by ATLAS for their $139 ~ \text{fb}^{-1}$ data set for different values of $\zeta_{12D}$. }
\end{figure}
\begin{equation}\label{eq:muHaZ}
	\mu_{\text{SM},{\cal{A}}} = \frac{ \Gamma(H)_{\text{SM}}}{ \Gamma( H \rightarrow \mathcal{A} \, Z) +  \Gamma(H)_{\text{SM}}}\,,
\end{equation}
with $\Gamma(H)_{\text{SM}}\simeq 4~\text{MeV}$ as the total Higgs boson decay width in the SM~\cite{LHCHiggsCrossSectionWorkingGroup:2011wcg}.
To constrain this BSM decay, we use the well constrained and hence representative signal strength for $H \rightarrow \gamma \gamma $. This has been measured 
\begin{equation}
\mu_{\gamma\gamma}=1.04^{+0.1}_{-0.09}~\hbox{\cite{ATLAS:2022tnm} }
\end{equation}
for the representative ATLAS Run 2 dataset of $139~\text{fb}^{-1}$.  For the on-shell decay of $H\to \mathcal{A}\,Z$, the maximum value of ALP mass allowed kinematically is $\simeq 34 ~\text{GeV}$ with $M_{H} = 125 ~\text{GeV}$ and $M_{Z} = 91.18 ~ \text{GeV}$. For heavier ALP masses, the branching ratio quickly dies off due to the offshellness of the involved $Z$ boson.

The allowed parameter space in $a_{2D}/f_{\mathcal{A}} $ vs $ M_{\mathcal{A}} $  plane  is shown in Fig.~\ref{fig:a2Dvsma} for three different values of $\zeta_{12D}$.  
The above $ 95\%$ limit translates into the lower bound on $\Gamma( H \rightarrow \mathcal{A} \, Z) < 0.65 ~ \text{MeV}$ using Eq.~\eqref{eq:muHaZ} for $\zeta_{12D}=1$.
The above bound on $\Gamma( H \rightarrow \mathcal{A} \, Z)$ is reduced by half with the HL-LHC projections for $H \rightarrow \gamma \gamma $ at $3~ \text{ab}^{-1}$ \cite{ATLAS:2018jlh}, i.e. we obtain $\Gamma( H \rightarrow \mathcal{A} \, Z) < 0.32 ~ \text{MeV}$ for $\zeta_{12D}=1$. 

\section{Higgs signals of Virtual ALPs}
\label{sec:virt}
\subsubsection*{Propagation vs. on-shell properties: Four-top production}
BSM corrections to the Higgs self energy $\Sigma_H$ can give rise to an oblique correction
\begin{equation}
	\hat{H} = - \frac{M_H^2}{2} \Sigma_H^{\prime \prime} (M_H^2)\;,
\end{equation}
analogously to the $\hat{W}$, $\hat{Y}$ parameters in the gauge sector, e.g.~\cite{Barbieri:2004qk}. Such a correction leads to a Higgs propagator modification~\cite{,Englert:2019zmt}
\begin{equation}
\label{eq:twopoint}
	-i\Delta_H(q^2) = \frac{1}{q^2 - M_H^2} \left( 1 + \hat{H} \left(1-{q^2\over M_H^2}\right) \right) \,,
\end{equation}
indicating a departure for large momentum transfers at unit pole residue.
Measurements of this parameter have by now been established by ATLAS and CMS in Refs.~\cite{ATLAS:2023ajo,CMS:2019rvj}. The expected upper limit is 
\begin{equation}
	\label{eq:hhatupperlim}
	\hat{H} \leq 0.12\,,
\end{equation}
at $95$\%~CL from the recent four-top production results of Ref.~\cite{ATLAS:2023ajo}. We can re-interpret this in the framework that we consider.  In parallel, we can employ an extrapolation of four-top final states to estimate sensitivity improvements that should become available in $\hat H$-specific analyses at the high-luminosity LHC (ATLAS currently observe a small tension in their $\hat H$ fit).

Explicit calculation in general $R_\xi$ gauge of the ALP insertion of Eq.~\eqref{eq:dec} into the Higgs two-point function yields the $\xi$-independent result (see also remarks in~\cite{Herrero:2020dtv,Herrero:2021iqt,Anisha:2022ctm})
\begin{multline}
\label{eq:2pointdiv}
\Gamma(H(q)H(q))=\frac{a_{2D}^2 \,  \zeta_{12D}^2}{4  \pi^2 f_{\mathcal{A}}^2}(4 M_{\mathcal{A}}^4 - 3 M_{\mathcal{A}}^2 \, q^2 \\+ 
		( q^2 - 3 M_{Z}^2 ) q^2) \Delta_{\text{UV}} + \dots\,,
\end{multline}
with $\overline{\text{MS}}$ factor
\begin{equation}
\label{eq:div}
\Delta_{\text{UV}}={\Gamma(1+\epsilon) \over \epsilon} \left({4\pi\mu^2\over M_H^2}\right)^\epsilon
\end{equation}
in dimensional regularisation $D=4-2\epsilon$ with `t Hooft mass $\mu$.  The ellipses in Eq.~\eqref{eq:2pointdiv} denote finite terms for $\epsilon\to 0$ (see below).

In the following we will adopt the on-shell scheme for field and mass renormalisation (cf.~Eq.~\eqref{eq:twopoint}), and the $\overline{\text{MS}}$ scheme for HEFT parameters (see also~\cite{Delgado:2013hxa,Herrero:2020dtv,Herrero:2021iqt,Asiain:2021lch,Anisha:2022ctm,Herrero:2022krh}).
On the one hand, part of the divergence of Eq.~\eqref{eq:2pointdiv} are then cancelled by the (divergent, div.) counterterms related to the Higgs wave function and mass renormalisation
\begin{alignat}{1}
	\delta Z_H\big|_{\text{div.}} &=  \frac{3 \, a_{2D}^2 \, \zeta_{12D}^2  (M_{\mathcal{A}}^2 + M_{Z}^2) }{4 \pi^2 f_{\mathcal{A}}^2 }\,,  \\
	\delta M_H^2 \big|_{\text{div.}} &= \frac{ a_{2D}^2  \, \zeta_{12D}^2  (4 M_{\mathcal{A}}^4 - 3 M_{\mathcal{A}}^2 M_H^2 - 3 M_{H}^2 M_{Z}^2) }{4\pi^2  f_{\mathcal{A}}^2 }\,. \nonumber
\end{alignat}
On the other hand, the appearance of a $~q^4$ contribution signifies the sourcing of the chiral dimension-4 operator $\mathcal{O}_{\square \square}$ of the HEFT Lagrangian
\begin{equation}
	\label{eq:opboxbox}
	\mathcal{O}_{\square \square}= a_{\square \square }  \frac{\square H \square H}{v^2}\,.
\end{equation}
This operator is renormalised by the ALP interactions via
\begin{equation}
 \delta a_{\square \square}= - \frac{	 a_{2D}^2  \zeta_{12D}^2 v^2}{8 \pi^2 f_{\mathcal{A}}^2 } \Delta_{\text{UV}}\,.
\end{equation}
Together, the renormalised Higgs two-point function then links to the $\hat H$ parameter as
\begin{widetext}
\begin{equation}
	\label{eq:hhatmatch}
	\begin{split}
\hat H =
- \frac{a_{2D}^2 M_{H}^2 \zeta_{12D}^2}{8  \pi^2 f_{\cal{A}}^2}
		\bigg( 2 B_{0}(M_H^2, M_{\cal{A}}^2, M_{Z}^2)\big|_{\text{fin.}} - 4(M_{\cal{A}}^2 - M_H^2 + M_Z^2)B'_{0}(M_H^2, M_{\cal{A}}^2, M_{Z}^2) \bigg. \\ \bigg.+ \left[M_{\cal{A}}^4 - 2 M_{\cal{A}}^2 (M_H^2 + M_Z^2) + (M_H^2 - M_Z^2)^2\right]  B''_{0}(M_H^2, M_{\cal{A}}^2, M_{Z}^2)\bigg) \,,
	\end{split}
\end{equation}
\end{widetext}
where `fin.' denotes the UV finite part of the Passarino-Veltman $B_0$ function after subtracting Eq.~\eqref{eq:div} and derivatives are taken with respect to the first argument of the $B_0$ function (an explicit representation can be found in Ref.~\cite{Denner:1991kt}). $\hat H$ vanishes in the decoupling limit $f_{\cal{A}}>M_{\cal{A}}\gg M_H$.

Equation~\eqref{eq:hhatmatch} shows that propagator corrections that can be attributed to $\hat H$ probe similar couplings as the Higgs decay of Eq.~\eqref{eq:dwhaz}, however, in a momentum transfer-enhanced way, at the price of a loop suppression. This way the energy coverage of the LHC that becomes under increasing statistical control provides additional sensitivity beyond the fixed scale Higgs decay. Any enhanced sensitivity to the on vs. off-shell phenomenology that can be gained from the combination of the processes discussed so far, can then break the degeneracies between the different HEFT coefficients in Eq.~\eqref{eq:alpflare}.

\begin{figure}[!t]
	\includegraphics[width=0.47\textwidth]{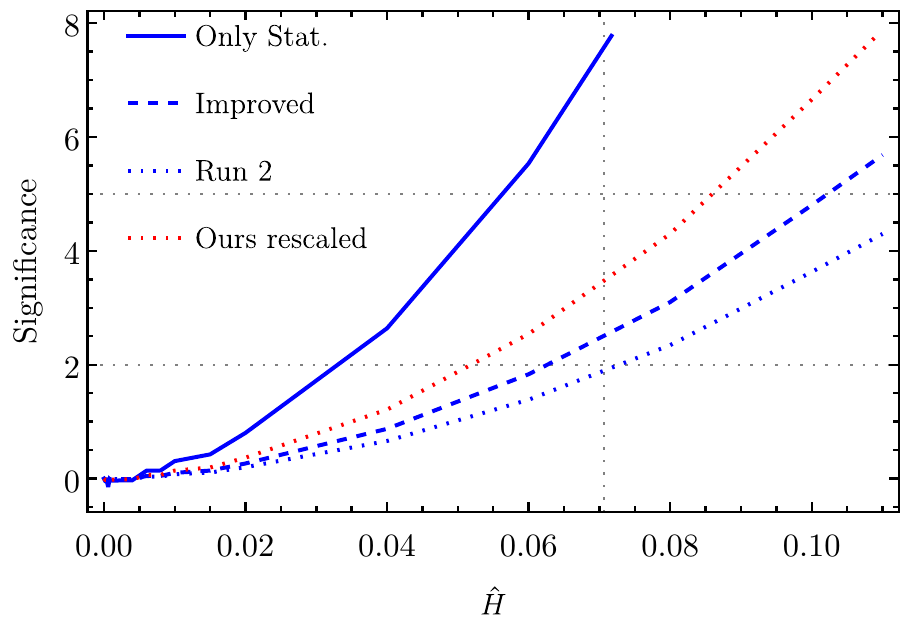}
	\caption{Expected significance for different values of $\hat{H}$ at HL-LHC using the projections of Ref.~\cite{CMS:2018nqq} (blue) and our estimates from projections using~Eq.~\eqref{eq:hhatupperlim} (red). \label{fig:hhatsign}}
\end{figure}

To obtain an extrapolation estimate from the current constraints on $\hat{H}$, we implement the modifications from $\hat{H}$ in {\tt{MadGraph5\_aMC@NLO}}~\cite{Alwall:2014hca} in order to estimate the changes caused in the four-top cross section from different contributions to the Higgs self-energy, and extrapolate the result of Eq.~\eqref{eq:hhatupperlim}. Assuming a significance $S(\hat{H} = 0.12) / \sqrt{B} = 2$ from the constraint of Ref.~\cite{ATLAS:2023ajo} at $140$/fb, and then subsequently rescaling the results to $3$/ab, we obtain the approximate significance at HL-LHC. While using the more recent results yields improved bounds compared to earlier projections of Ref.~\cite{CMS:2018nqq} that include systematics (due to improvements in the analysis procedure utilising ML techniques), our projections remain conservative compared to the previously estimated significance with only statistical uncertainties, see Fig.~\ref{fig:hhatsign}. 

\begin{figure}[!t]
	\includegraphics[width=0.47\textwidth]{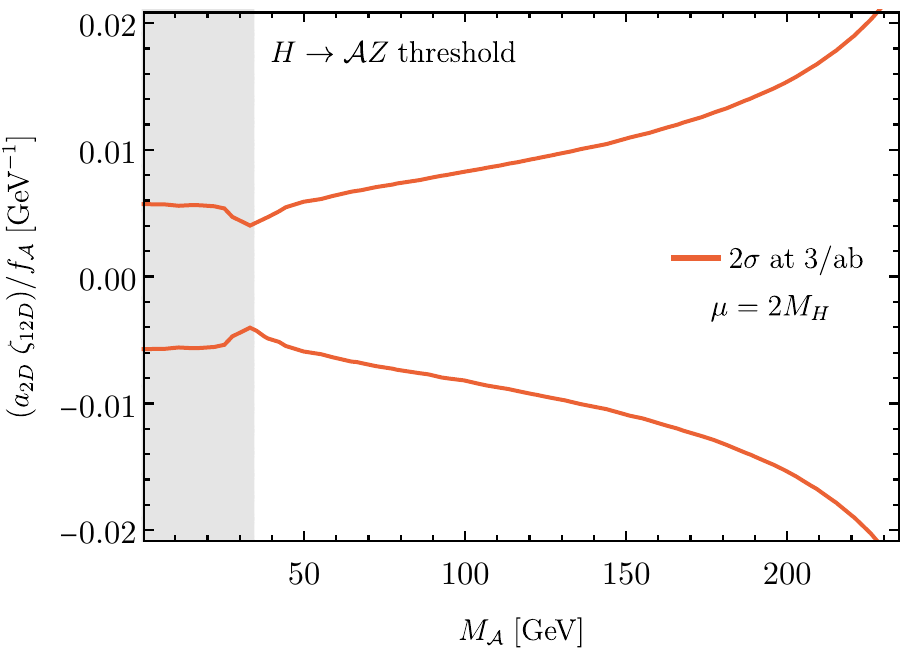}
	\caption{The solid line indicates the $2 \sigma$ contour on the $a_{2D}/f_{\cal{A}}$-$M_{\cal{A}}$ plane from the $\hat{H}$ analysis for a scale $\mu = 2 M_H$. The shaded region indicates the open $H \rightarrow {\cal{A}} Z$ decay channel. \label{fig:fourtopbands}}
\end{figure}

In Fig.~\ref{fig:fourtopbands}, we also see that if $M_{\cal{A}}$ is light, it will freely propagate in the 2 point function thus imparting the characteristic $q^4$ dependence probed by $\hat H$. This also means that this behaviour is essentially independent of the light ALP mass scale. Turning to heavier states, this kinematic dependence is not sourced as efficiently anymore, leading to a quick decoupling from the two-point Higgs function and reduced sensitivity and larger theoretical uncertainty. We will return to the relevance of $\hat H$ for the discussed scenario after discussing the modifications to Higgs pair production in the next section.

\subsubsection*{Higher terms of the ALP flare function: Higgs pair production}
Corrections to Higgs pair production under the same assumptions as in the previous section are contained in propagator corrections and corrections to trilinear Higgs coupling. As with the chiral dimension-4 operator that leads to new contributions to the Higgs-two point function, there are additional operators that modify the Higgs trilinear interactions. The amputated off-shell three-point function receives contributions (see also~\cite{Herrero:2022krh})
\begin{multline}
v^3\Gamma_1(H(q)H(k_1)H(k_2)) = 
 a_{\chi 1} (q^4 + k_1^4 + k_2^4)  \\
+  2 a_{\chi 2} (q^2 k_1^2 + k_1^2 k_2^2 + q^2 k_2^2)
+  a_{\chi 3}v^2 (q^2+ k_1^2 + k_2^2)\,,
\end{multline}
which are renormalised in the $\overline{\text{MS}}$ scheme according to
\begin{equation}
\begin{split}
\delta a_{\chi1} &= \frac{a_{2D}^2  \zeta_{12D} v^2}{8 \pi^2 f_{\mathcal{A}}^2 }  (3(1+\zeta_1) \zeta_{12D} + 2\zeta_{22D}) \Delta_{\text{UV}}\,,\\
\delta a_{\chi2} &= \frac{ 3 a_{2D}^2 \zeta_{12D}^2 v^2 }{8 \pi^2 f_{\mathcal{A}}^2  } (1+\zeta_1) \Delta_{\text{UV}} \,,\\
\delta a_{\chi3} &= \frac{ 3 a_{2D}^2 \zeta_{12D} v^2}{4 \pi^2 f_{\mathcal{A}}^2  } \big[
(M_{\cal{A}}^2+M_Z^2)\zeta_{22D} \\ & \hspace{1.2cm} - 3(M_{\cal{A}}^2+2M_Z^2) (1+\zeta_1)\zeta_{12D}\big]\Delta_{\text{UV}}\,.
\end{split}
\end{equation}
The remaining renormalisation of the chiral dimension-2 term follows from Eq.~\eqref{eq:hpot}
\begin{multline}
\delta\Gamma_2(H(q)H(k_1)H(k_2))\big|_\text{div} = -\frac{ 9a_{2D}^2 \zeta_{12D}^2 }{8 \pi^2 f_{\mathcal{A}}^2 }  \kappa_3 (M_{\cal{A}}^2+m_Z^2)   \\
-\frac{ a_{2D}^2 \zeta_{12D} M_{\cal{A}}^4 }{2 \pi^2 f_{\mathcal{A}}^2 v } \left( 2 (1+\zeta_1) \zeta_{12D} - \zeta_{22D} \right)\,. 
\end{multline}
\begin{figure*}[!t]
	\subfigure[\label{fig:histo}]{\includegraphics[width=0.47\textwidth]{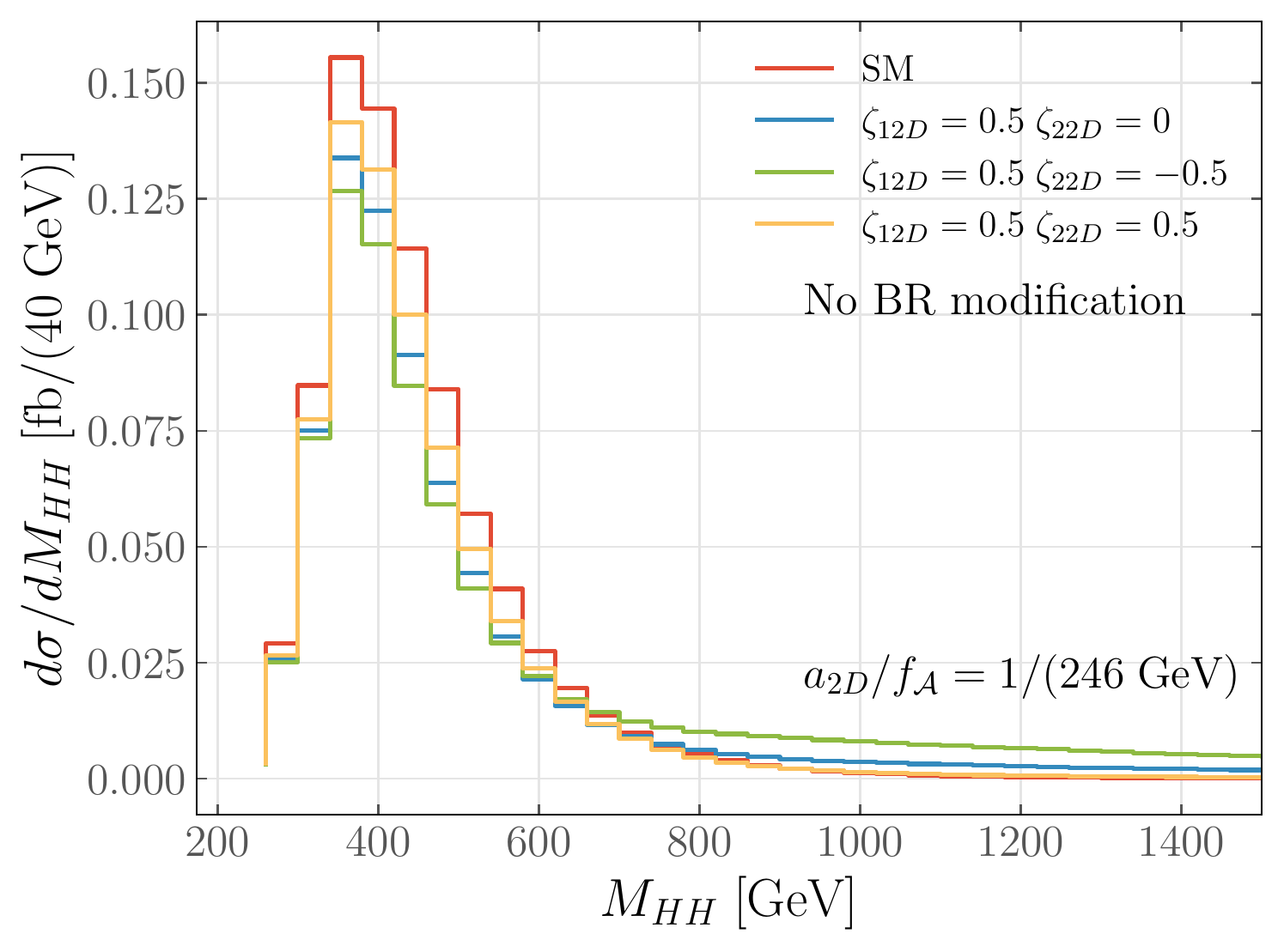}}
	\hfill
	\subfigure[\label{fig:regio}]{\includegraphics[width=0.47\textwidth]{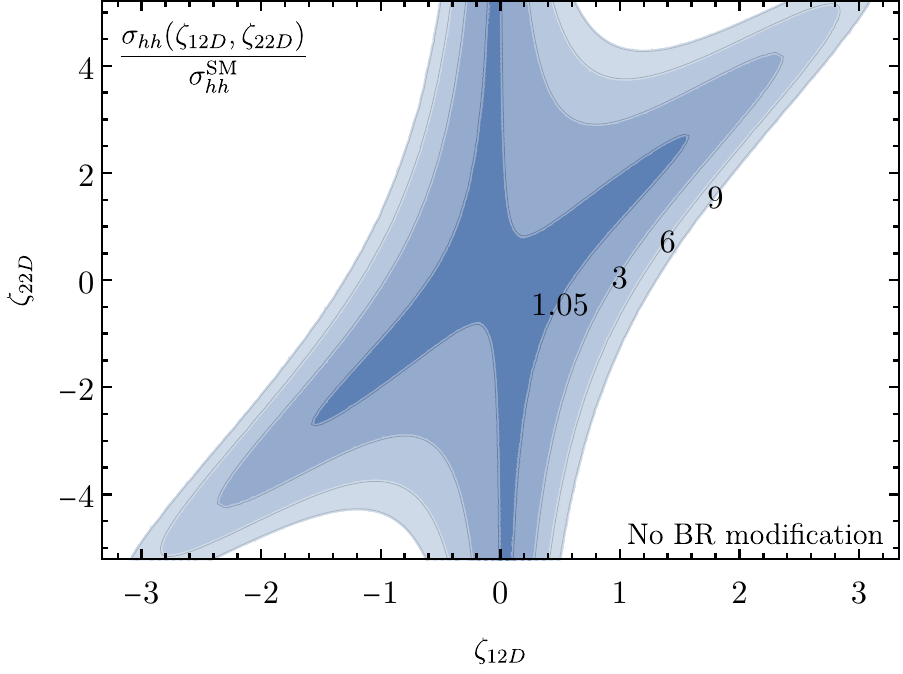}}
	\caption{\label{fig:invm} Invariant mass of the Higgs pair $M^2_{HH}=(p_{H,1}+p_{H,2})^2$ for different values of $\zeta_{12D,22D}$ and the SM in $gg\to HH$, and $M_{\cal{A}}=1~\text{GeV}$ is shown on the left. Considering the final state Higgs boson on-shell, the corrections to their decays according to Eq.~\eqref{eq:dwhaz} are not included. The normalisation is chose to $\sigma(HH)^\text{SM}\simeq 32~\text{fb}$~\cite{LHCHiggsCrossSectionWorkingGroup:2016ypw} and we use SM K factors to qualitatively include higher order QCD corrections. Squared contributions from the renormalised $gg\to hh$ amplitude are included. On the right the ratio of the new-physics cross section with respect to the SM is shown for different values in the $\zeta_{12D}$-$\zeta_{22D}$ plane.}
\end{figure*}
\begin{figure*}[!t]
	\subfigure[\label{fig:light}]{\includegraphics[width=0.47\textwidth]{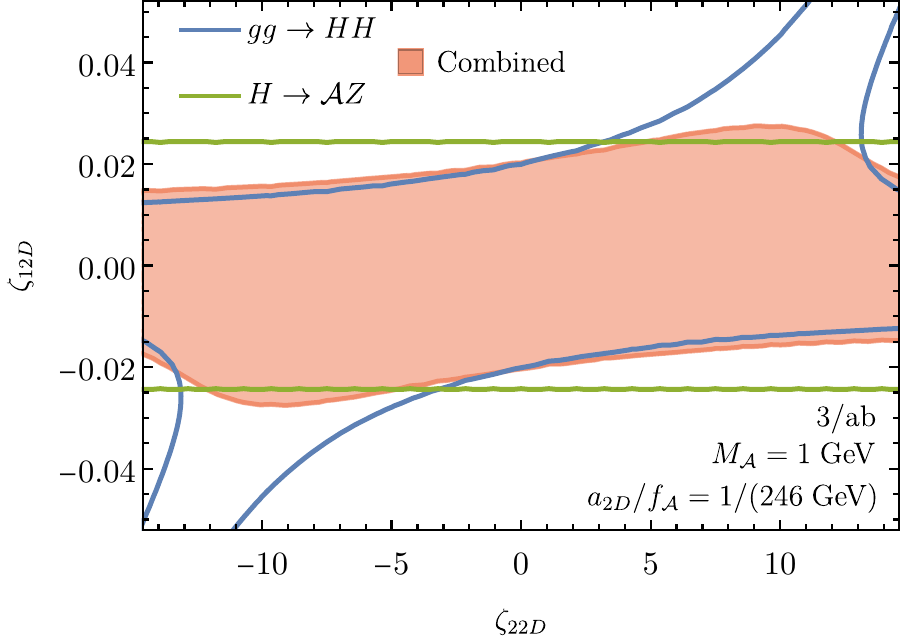}}
	\hfill
	\subfigure[\label{fig:heavy}]{\includegraphics[width=0.47\textwidth]{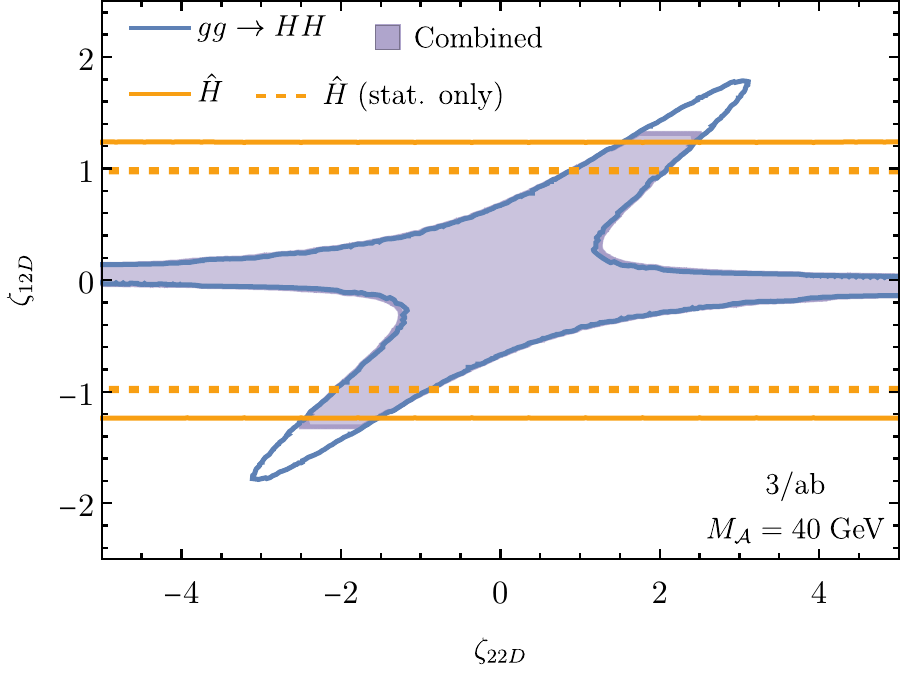}}
	\caption{\label{fig:comb_contours} $95$\% CL regions for the case of a light $M_{\cal{A}}=1~\text{GeV}$ (left) and heavy $M_{\cal{A}}=40~\text{GeV}$ (right) ALP.}
\end{figure*}

ATLAS (CMS) have set highly competitive expected 95\% confidence level cross section limits of $\sigma/\sigma_{\text{SM}}<3.9~(5.2)$~\cite{CMS:2022hgz,ATLAS:2022xzm} in the $b\bar b \tau \tau$ channel~\cite{Dolan:2012rv} alone. Slightly reduced sensitivity~\cite{ATLAS:2021ifb,CMS:2022cpr,CMS:2020tkr} can be achieved in the $4b$ and $2b2\gamma$ modes~\cite{Baur:2003gp,FerreiradeLima:2014qkf}. ATLAS have combined these channels to obtain a combined exclusion of $3.1 \sigma_{\text{SM}}$~\cite{ATLAS:2021tyg} with the currently available data and forecast a sensitivity of $\sigma/\sigma_{\text{SM}}\gtrsim 1.1$ at the HL-LHC~\cite{ATLAS:2022faz}. We use the two latter result to gain a qualitative sensitivity reach of Higgs pair production in the considered scenario.  

In Fig.~\ref{fig:invm}, we show representative invariant Higgs pair mass distributions for 13 TeV LHC collisions, which demonstrates the potential of multi-Higgs final states' sensitivity to the momentum-enhanced
new physics contributions characteristic to the ALP.\footnote{We have implemented these changes into an in-house Monte Carlo event generate based on~{\sc{Vbfnlo}}~\cite{Arnold:2008rz,Baglio:2011juf} employing {\sc{FeynArts}}, {\sc{FormCalc}}, and {\sc{LoopTools}}~\cite{Hahn:1998yk,Hahn:2000jm,Hahn:2000kx,Hahn:2006qw} and {\sc{PackageX}}~\cite{Patel:2016fam} for numerical and analytical cross checks. Throughout this work we chose a renormalisation scale of $\mu=2M_h$.}
 The behaviour exhibited by the invariant mass distribution is not sensitive to the mass of the ALP as long as the latter is not close to the $\simeq 2M_H$ threshold that determines the $gg\to HH$ phenomenology. In instances when hefty ALPs propagate freely, their distinctive momentum enhancements will sculpt the Higgs-boson distributions. In parallel, non-linear effects will be important away from the SM reference point as shown in Fig.~\ref{fig:invm}. This shows that the constraints that can be obtained in the di-Higgs channel are relatively strongly coupled, which is motivation for us to directly include ``squared'' $\text{BSM}$ effects to our analysis in addition to interference effects.
  
We combine the three representative analyses in a global $\chi^2$ to obtain sensitivity estimates. In the case when the ALP is light, there are significant modifications to Higgs physics, also at large momentum transfers, see also Fig.~\ref{fig:invm}. Of course, these large contributions in particular to the Higgs pair rate are tamed by decreasing signal strengths into SM-like states, which quickly result in tension with experimental observations for larger couplings. As Higgs pair production observations need to rely on relatively clean and high branching ratio final states, the prospects of Higgs pair production (and four top) analyses to provide additional sensitivity is relatively low. This is highlighted already in the combination of the Higgs decay constraints with these processes in Fig.~\ref{fig:light}. 

For parameter choices for which the ALP is above the Higgs decay threshold, this picture changes. Multi-Higgs constraints remain relatively insensitive to the ALP mass scale as long as these states are away from the $2 M_H$ threshold. The cross section enhancement then translates directly into an enhancement of the observable Higgs boson pair production rate. In turn, constraints on the the higher order terms in the ALP flare function become possible. It is important to note that these are independent of couplings (to first order) that shape the ALP decay phenomenology. 

As the large enhancements result from the tails of distributions there is a question of validity. Nonetheless the momentum dependence introduced by Eq.~\eqref{eq:dec} leads to partial wave unitarity violation as, e.g. $HZ$ scattering proceeds momentum-enhanced. A numerical investigation shows that for ${\cal{O}}(1)$ couplings in Eq.~\eqref{eq:axlag}, conserved zeroth partial wave unitarity up to scales $\sim 1.5~\text{TeV}$ sets a lower bound of $f_a\gtrsim 300~\text{GeV}$ for unsuppressed propagation ${M}_{\cal{A}}=1~\text{GeV}$. These constraints are driven by the longitudinal $Z$ polarisations, constraints from transverse modes are comparably weaker. This means that the entire region that is shown in Fig.~\ref{fig:histo} is perturbative at tree-level. In parallel, the HL-LHC is unlikely to probe Higgs pairs beyond invariant masses $M_{HH}>600~\text{GeV}$ in the SM (for which the cross section drops to 10\% of the inclusive rate). Most sensitivity in HL-LHC searches results from the threshold region. Therefore, the sensitivity expected by the HL extrapolation of \cite{ATLAS:2022faz} will probe Eq.~\eqref{eq:axlag} in a perturbatively meaningful regime.

The combined constraints are largely driven by Higgs pair constraints, Fig.~\ref{fig:heavy}. However, it is worth highlighting that the statistics-only extrapolation does not include changes to the four top search methodology. Improvements of the latter can be expected with increasing luminosity and the final verdict from four top production might indeed be much more optimistic than our $\sqrt{\text{luminosity}}$ extrapolation might suggest.
\section{Summary and Conclusions}
\label{sec:conc}
Searches for new light propagating degrees of freedom such as axion-like particles are cornerstones of the BSM programme in particle physics as explored at, e.g., the Large Hadron Collider.
The Higgs boson, since a global picture of its interactions is still incomplete, provides a motivated avenue for the potential discovery of new physics in the near future as the LHC experiments gain increasingly phenomenological sensitivity in rare processes that could be tell-tale signs of Higgs-related BSM physics. 

We take recent experimental developments in multi-Higgs and multi-top analyses as motivation to analyse effective Higgs-philic ALP interactions, also beyond leading order. This enables us to tension constraints from different areas of precision Higgs phenomenology, combining Higgs decay modifications with large-momentum transfer processes that are becoming increasingly accessible at the LHC. For light states and sizeable HEFT-like couplings, a large part of the sensitivity is contained in Higgs signal strength measurements (see also~\cite{Brivio:2017ije}), which, however, only provide limited insights into the Higgs-ALP interactions. Higher terms of the Higgs-ALP flare function, still have the phenomenological potential to sizeably modify Higgs pair final states at a level that will be observable at the LHC in the near future.
Our findings therefore also highlight further the relevance of multi-top and multi-Higgs final state for the quest for new physics.

\acknowledgements
We thank Dave Sutherland for insightful discussions. C.E. thanks the high-energy physics group (FTAE) at the University of Granada for their hospitality during early stages of this work.

A.~is supported by the Leverhulme Trust under grant RPG-2021-031. S.D.B is supported by SRA (Spain) under Grant No.\ PID2019-106087GB-C21 (10.13039/501100011033), and PID2021-128396NB-100/AEI/10.13039/501100011033;
by the Junta de Andalucía (Spain) under Grants No.~FQM-101, A-FQM-467-UGR18, and P18-FR-4314 (FEDER). C.E. is supported by the STFC under grant ST/T000945/1, the Leverhulme Trust under grant RPG-2021-031, and the Institute for Particle Physics Phenomenology Associateship Scheme. 
P.S. is supported by the Deutsche Forschungsgemeinschaft under Germany’s Excellence strategy EXC2121
``Quantum Universe'' - 390833306. This work has been partially funded by the Deutsche Forschungsgemeinschaft (DFG, German Research Foundation) - 491245950.
%

\bibliography{paper.bbl}
\end{document}